\documentclass[aip,cha,preprint,numerical]{revtex4-2}
\usepackage{amsfonts,amssymb,amsmath,mathrsfs}
\usepackage{mathptmx}
\usepackage{bm}
\usepackage{xcolor}
\usepackage{graphicx}
\usepackage{dcolumn}
\usepackage{bm}
\usepackage{natbib}		
\usepackage{color}

\usepackage[T2A]{fontenc}
\usepackage[utf8]{inputenc} % включаем свою кодировку: koi8-r или utf8 в UNIX, cp1251 в Windows

\begin{document}
%---------------------------------------------------------------------------%
\title{Chimera states in a system of stationary and flying-through deterministic particles with an internal degree of freedom}

	\author{Maxim~I.~Bolotov}
    \email{maksim.bolotov@itmm.unn.ru}
	\affiliation{Department of Control Theory, Scientific and Educational Mathematical Center ``Mathematics of Future Technologies'', Nizhny Novgorod State University, Gagarin Av. 23, Nizhny Novgorod, 603950 Russia}

	\author{Lev~A.~Smirnov}
	\email{lev.smirnov@itmm.unn.ru}
	\affiliation{Department of Control Theory, Scientific and Educational Mathematical Center ``Mathematics of Future Technologies'', Nizhny Novgorod State University, Gagarin Av. 23, Nizhny Novgorod, 603950 Russia}

	\author{Vyacheslav~O.~Munyaev}
    \email{munyaev@itmm.unn.ru}
	\affiliation{Department of Control Theory, Scientific and Educational Mathematical Center ``Mathematics of Future Technologies'', Nizhny Novgorod State University, Gagarin Av. 23, Nizhny Novgorod, 603950 Russia}
		
    \author{Grigory~V.~Osipov}
	\email{osipov@vmk.unn.ru}	\affiliation{Department of Control Theory, Scientific and Educational Mathematical Center ``Mathematics of Future Technologies'', Nizhny Novgorod State University, Gagarin Av. 23, Nizhny Novgorod, 603950 Russia}
\begin{abstract}
    We consider the effect of the emergence of chimera states in a system of coexisting stationary and flying-through in potential particles with an internal degree of freedom determined by the phase. All particles tend to an equilibrium state with a small number of potential wells, which leads to the emergence of a stationary chimera. An increase in the number of potential wells leads to the emergence of particles flying-through along the medium, the phases of which form a moving chimera. Further, these two structures coexist and interact with each other. In this case, an increase in the local synchronization degree of the chimera is observed in the areas of the synchronous cluster location.
\end{abstract}
\date{\today}
\pacs{05.45.Xt, 45.20.dc}
\date{\today}
\maketitle
	\begin{quotation}
        Chimera states are one of the key spatiotemporal structures that are realized in systems of nonlocally coupled identical oscillators with attractive interaction. Moreover, they coexist with a completely synchronous state, characteristic of systems with spatial symmetry. However, particle motion can destroy the fully synchronous state, leading to the dominance of the chimera regime. Here we show that when placed in a periodic potential, non-local interactions can cause particles to split into two subsystems of stationary and flying-through elements. At the same time, within each of them, coexisting chimera states are formed.
	\end{quotation}
\maketitle
%---------------------------------------------------------------------------%
\section{Introduction}\label{sec:Introduction}
{The study of active particles is a current research topic at the intersection of physics, chemistry, biology and engineering~\cite{Reimann2002,Hanggi2009,Reichhardt2017}. From a physical point of view, active particles and the structures they form represent a new complex object of research, located at the intersection of statistical, chemical and biological physics, in which many nonequilibrium and nonlinear effects are observed.~\cite{Bechinger2016}. Active particles are usually considered to be internally fluctuating elements, so that most effects are observed in the presence of noise~\cite{lowen2020}. In many cases, internal noise plays a fundamental role in the behavior of active particles. The influence of noise is strongly pronounced for microscopic objects located in a fluctuating medium, for example, in active Brownian particles. Furthermore, internal fluctuations are a significant factor when modeling macroscopic objects such as animals~\cite{Vicsek2012}, transport~\cite{kerner2019} and pedestrians~\cite{helbing2013}. However, in the case of the ``superactivity limit'', when the self-motion is so intense that the speed of the particles can be considered practically constant, a purely deterministic description of active particles is possible~\cite{aranson2022}. Purely deterministic dynamics have been studied much less~\cite{ebeling1999, kruk2018}, perhaps because the initial studies only observed rather trivial periodic regimes~\cite{ebeling1999}. Recently, interest in deterministic dynamics has revived in the context of the study of the active particles movement in a disordered environment~\cite{peruani2018}.
}

{
Active particles are also an adequate model for describing colloidal systems~\cite{Ebbens2016, Dietrich2018, Fehlinger2023, Cereceda2024}. These systems are of interest for applications involving controlled transport and delivery of chemical or biological payloads attached to functionalized particles or for high-precision particle sorting and fractionation.
}

{
Active particles can also have their own dynamics. In this case, the state of each particle can be described, for example, by a phase~\cite{igoshin2001}. If the phase dynamics of particles affects the position of a particle during non-local spatial interaction, the particle is called a swarmalator~\cite{keeffe2017}. Such models make it possible to describe colloidal suspensions of magnetic particles and other biological and physical systems in which the effects of self-assembly and synchronization interact~\cite{snezhko2011, yan2012, keeffe2019}.
}

{
The main effects of phase dynamics are complete synchronization, cluster states, splay states, and chimera states. The chimera is a state in which an ensemble of identical elements is divided into groups of fully synchronous oscillators and groups of asynchronous oscillators~\cite{Abrams2004}. Such regimes have been found in both lumped~\cite{mishra2023}, and distributed systems~\cite{omel2018, parastesh2021}.
In distributed systems with non-local coupling, such regimes can be observed in the region of attractive interactions and coexist with fully synchronous regimes. In~\cite{smirnov2021} it was shown that the motion of oscillators can lead to the destruction of the stability of the fully synchronous regime. In this case, the chimera state became the dominant dynamical mode.
In~\cite{wang2019}, it was demonstrated that oscillators mobility can act as noise and cause jumps from metastable dynamical states to a globally synchronized state for attractive coupling. In the case of repulsion, this breaks the chimera state stability and further enhances the heterogeneity. In~\cite{petrungaro2019} it is shown that for a system of moving oscillators with a time delay in coupling, mobility can speed up synchronization.
}

{
In this paper, we consider a system of particles with mass and an internal degree of freedom described by a phase. The particles are located in a periodic potential and are subject to a constant force and dissipation. In this work, we assume that phase dynamics does not affect particle coordinates.
Section~\ref{sec:model} describes the model and its spatial and phase dynamics. The influence of the potential wells number on the stationary and flying-through particles number is described. Section~\ref{sec:stat} describes chimera states in a system of stationary particles, when several oscillators can be in a potential well. Section~\ref{sec:dyn} describes the dynamics in a system of coexisting stationary and flying-through particles, when chimera states in subensembles coexist with each other. Section~\ref{sec:Conclusion} provides a brief description of the results obtained.
}
\clearpage
%---------------------------------------------------------------------------%
\section{Model}~\label{sec:model}
\subsection{Particle dynamics}
Let us consider a system of $N$ particles with masses $\mu$, located on a segment of length $L$ with periodic boundary conditions (i.e. on a ring). The position of each particle is characterized by a coordinate $x_n \in [0, L)$ ($n=1, 2, \dots, N$). Particles on the ring move in a potential given by the function $U=U(x_n)$, under the action of a constant force $\gamma$. It is assumed that the system also has a friction force with a coefficient $\lambda$. Then the dynamics of the particle coordinates is determined by the following equation
\begin{equation}
    \mu \ddot{x}_n + \lambda \dot{x}_n = f(x_n) + \gamma,
    \label{eq:orig_x}
\end{equation}
where $f(x_n) = -dU/dx_n$ defines the potential force acting on the particle. Let us dwell on the case of a periodic potential with $Q$ minima equidistantly located on the ring
\begin{equation}
    U(x_n) = -L \cos\big(2\pi Q x_n/L\big)\big/2\pi Q.
    \label{eq:pot}
\end{equation}
The system \eqref{eq:orig_x} at \eqref{eq:pot} takes the following form
\begin{equation}
    \mu \ddot{x}_n + \lambda \dot{x}_n + \sin\big(2\pi Q x_n/L\big) = \gamma.
    \label{eq:main_x}
\end{equation}

\begin{figure}
    \centering
\includegraphics[width=0.70\columnwidth]{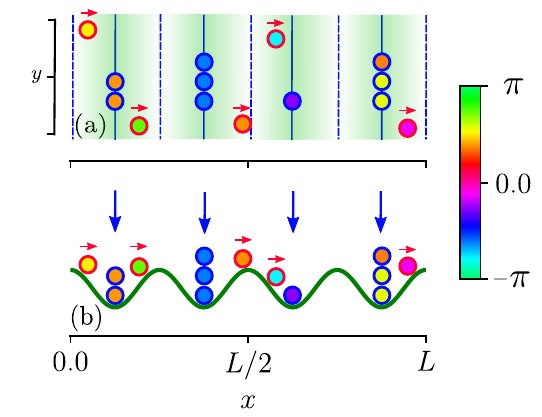}
    \caption{{Schematic illustration of a system of particles defined by the equations~\eqref{eq:main_x}, \eqref{eq:main_phi}, \eqref{eq:ker}, in the quasi-one-dimensional (a) and one-dimensional (b) cases. The green curve corresponds to the potential function \eqref{eq:pot} with $Q=4$. Blue solid lines (blue arrows) indicate the position of potential wells, blue dotted lines indicate the position of potential maxima, red arrows indicate the direction of movement of flying-through particles. Markers with blue (red) borders are stationary (flying-through) particles. The fill color of the markers corresponds to the phase $\varphi_n$ of the corresponding particle. In one potential well there can be several particles, the phases of which may differ.}}
    \label{fig:0}
\end{figure}

{Equation~\eqref{eq:main_x} also describe the evolution of particles in a quasi-one-dimensional medium, where the distribution of particles along the second coordinate $y$ makes the probability of their collision sufficiently small, while the dynamics of the coordinate $y$ is sufficiently slow, which allows us to consider only the coordinate $x$ (see fig.~\ref{fig:0}).}

It is easy to show that the equation~\eqref{eq:main_x} by replacing the coordinate $x_n=X_n L\big/2\pi Q$ and time $t=\tau\sqrt{\mu L\big/2\pi Q} $ can be reduced to the pendulum equation~\cite{andronov2013}
\begin{equation}
    X_n^{''} + \bar{\lambda}X_n^{'} + \sin X_n = \gamma,
    \label{eq:pendula}
\end{equation}
where
\begin{equation}
    \bar{\lambda} = \lambda\sqrt{L\big/2\pi \mu Q}
    \label{eq:lambda}
\end{equation}
-- linear coefficient of effective dissipation.

\begin{figure}
    \centering
\includegraphics[width=0.50\columnwidth]{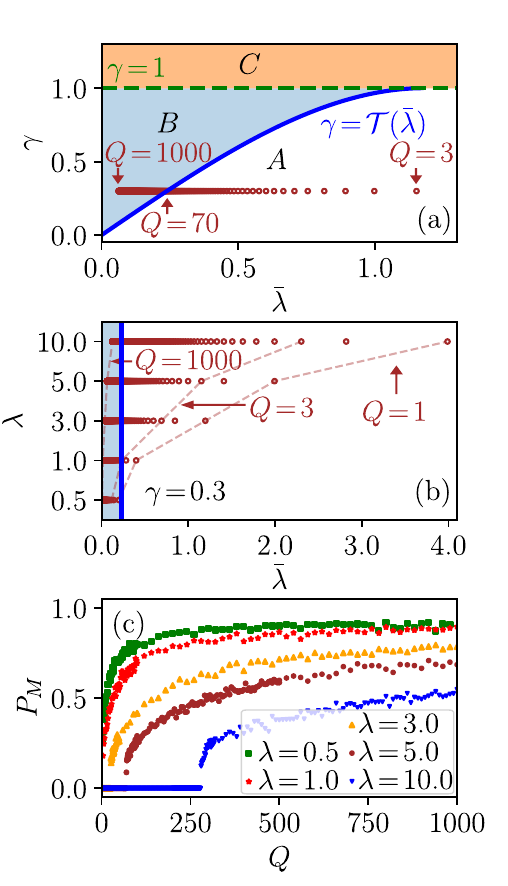}
    \caption{(a) The plane of parameters $\bar{\lambda}$, ${\gamma}$ of equation~\eqref{eq:pendula}. $\gamma=1.0$ (green dashed line), $\gamma = \mathcal{T}(\bar{\lambda})$ (blue curve) is the Tricomi curve. Brown dots are the values of the parameters corresponding to the parameters of the system~\eqref{eq:main_x}, at $\gamma=0.3$ for different values of $Q$. The value $Q=70$ corresponds to the transition from region $B$ to $C$. (b) Values of the system parameters~\eqref{eq:pendula} corresponding to different numbers of potential wells $Q$ in the system~\eqref{eq:main_x} at $\gamma=0.3$ for different values of the parameter $\lambda$ (the scale of the ordinate axis is nonlinear). Brown dotted lines connect the points corresponding to $Q=1$, $Q=3$ and $Q=1000$. (c) Probability $P_M$ of entering periodic rotational motion in the system~\eqref{eq:main_x} for $\gamma=0.3$ and different $\lambda$ depending on $Q$. Each point is calculated based on $100$ experiments. Initial values: the initial coordinate $x_n(0)$ was chosen from a uniform distribution on the interval $[0, L]$, the initial velocity was chosen from a normal distribution $N(v_0, \sigma^2)$ with $v_0=0.0$ and $\sigma = 0.1/\sqrt{\lambda}$. Parameters: $L=1.0$, $\mu=1.0$, $\lambda=5.0$.}
    \label{fig:1}
\end{figure}

The pendulum equation \eqref{eq:pendula} is well studied~\cite{andronov2013}. It is known that, depending on the parameters $\bar{\lambda}$ and $\gamma$, there may be one stable equilibrium state (region $A$ in Fig. \ref{fig:1}a), a stable equilibrium state and rotational periodic motion may coexist (region $B$ in Fig. \ref{fig:1}a), or only stable rotational motion may take place (region $C$ in Fig. \ref{fig:1}a). Based on this, each of the system equations~\eqref{eq:main_x}, depending on the values of the control parameters, can have $Q$ stable equilibrium states $x = x_q$ ($q=1,2, \dots, Q$), corresponding to local minima of the potential $U=U(x_n)$. There may also be a stable periodic rotational motion. If the coordinates $(x_n, \dot{x}_n)$ of any particle are in the attraction region of a stable equilibrium state, then after the transition process the position of the particle becomes stationary. A set of such particles is called \textit{stationary}. In the case of a periodic motion, the particle begins to rotate in a ring. We call a set of such particles \textit{flying-through}. Due to the existence of a region of the equilibrium state bistability  and rotational motion in the system \eqref{eq:pendula} (region $B$ in Fig. \ref{fig:1}a), depending on the initial conditions, the particles \eqref{eq:main_x} can be divided into two groups: stationary and flying-through.
A schematic representation of the system under consideration is shown in the Fig.~\ref{fig:0}.

\subsection{Particle phase dynamics}
Each particle is also characterized by an additional internal degree of freedom, defined by a cyclic variable -- the phase $\varphi_n \in (-\pi, \pi]$. The dynamics of the $\varphi_n$ phases is determined by the Kuramoto--Battogtokh model, which describes the non-local interaction between particles and is governed by the following system of equations
\begin{equation}
    \dot{\varphi}_n = \omega + \sum_{\Tilde{n}=1}^{N}G\big(x_{\Tilde{n}}-x_n\big)\sin\big(\varphi_{\Tilde{n}}-\varphi_n-\alpha\big),
    \label{eq:main_phi}
\end{equation}
where $\omega$ is the natural frequency, $\alpha$ is the phase shift parameter that determines the type of interaction between elements, $G(x)$ is the kernel that determines the nonlocal interaction. As the interaction kernel function we choose
\begin{equation}
    G(x) = \kappa \cosh \big( \kappa (|x| - L/2) \big) \Big/ 2 \sinh(\kappa L/2)
    \label{eq:ker}
\end{equation}
that well approximates nonlocal interaction. The quantity $\kappa$ denotes the scaling parameter of the kernel $G(x)$.
{Note that in our setup, phase dynamics does not affect the dynamics of particle coordinates.}

The dynamics of the system \eqref{eq:main_phi}, \eqref{eq:ker} for equidistant particles on a ring and for the continuum limit (infinite number of particles) has been well studied~\cite{kuramoto2002, Wolfrum2011, smirnov2017, bolotov2020}. This system is the first to demonstrate the possibility of realizing long-lived chimera states of different types that coexist with a fully synchronous state. {In~\cite{Wolfrum2011} it is shown that in the case of a finite number of oscillators the chimera state is a chaotic transient regime, but at the same time long-lived. Due to the attractive type of interaction, the fully synchronous mode is stable.} In the case of non-equidistant arrangement of particles or their motion according to a random law along the medium, it was shown that a completely synchronous state can be destroyed and evolve to a chimera state, which becomes dominant in the system of phase oscillators~\cite{smirnov2021}. {Thus, the chimera state is the most typical variant of phase dynamics demonstrated by a large group of stationary or moving particles along the medium.}

\subsection{Coexistence of stationary and flying-through particles}
In the framework of our study, we will focus on the value of control parameters~\eqref{eq:main_x}, at which the coexistence of stationary and flying-through particles is possible. Let us fix the number of elements $N=512$, the length of the medium $L=1.0$, the mass $\mu=1.0$ and the dissipation parameter $\lambda = 5.0$. This set of parameters is remarkable in that for $0<\gamma<1$ the nature of the particle dynamics~\eqref{eq:main_x} will significantly depend on the potential wells' number $Q$. For $Q\leq Q^{*}(\gamma, \lambda, \mu, L)$, where $Q^{*}(\gamma, \lambda, \mu, L)$ -- is some critical value, the particles will become stationary under any initial conditions. At $Q>Q^{*}$ a situation of bistability is possible when a part of particles is in potential wells and a part makes rotational motions along the medium. This is explained by the location of the corresponding pair $\bar{\lambda}$ and $\gamma$ on the parameter plane of the system~\eqref{eq:pendula} (see Fig.~\ref{fig:1}a, brown markers). Figure~\ref{fig:1}b shows the relationship between the parameter $\lambda$ of the system~\eqref{eq:main_x} and $\bar{\lambda}$ of the system~\eqref{eq:pendula} at different $Q$ and a fixed $\gamma=0.3$. It is seen from~\eqref{eq:lambda} that for the sufficiently small $\lambda$, the one-well potential with $Q=1$ (e.g., $\lambda=0.5$ in Fig.~\ref{fig:1}b) can already correspond to the region $B$ of the~\eqref{eq:pendula} system, where  the bistability occurs. For sufficiently large $\lambda$, the transition to the bistability region of the pendulum equation~\eqref{eq:pendula} occurs only at sufficiently large $Q$ (e.g., $\lambda=10.0$ in Fig.~\ref{fig:1}b).

Figure~\ref{fig:1}c shows the probability $P_M$ of the particle's transition to rotational motion as a function of the potential wells' number $Q$ for different values of the dissipation parameter $\lambda$ and $\gamma=0.3$. The initial coordinates $x_n(0)$ were sampled from uniform distribution on the segment $[0, L]$; the initial velocities $\dot{x}_n(0)$ were sampled from normal distribution $N(v_0, \sigma^2)$ with $v_0=0.0$ and $\sigma = 0.1/\sqrt{\lambda}$. It is easy to show that for $Q \to + \infty$, $P_M\to 1$, since $\bar{\lambda} \to 0$ (see~\eqref{eq:lambda}), which corresponds to the conservative version of the pendulum equation~\eqref{eq:pendula}, where the attraction region of the stationary state tends to zero. It is clear from Fig.~\ref{fig:1}c that when the number of potential wells $Q$ passes the critical value $Q^*$, the fraction of flying-through particles in the system starts to increase rapidly.

\begin{figure}[h!]
    \centering
    \includegraphics[width=0.8\columnwidth]{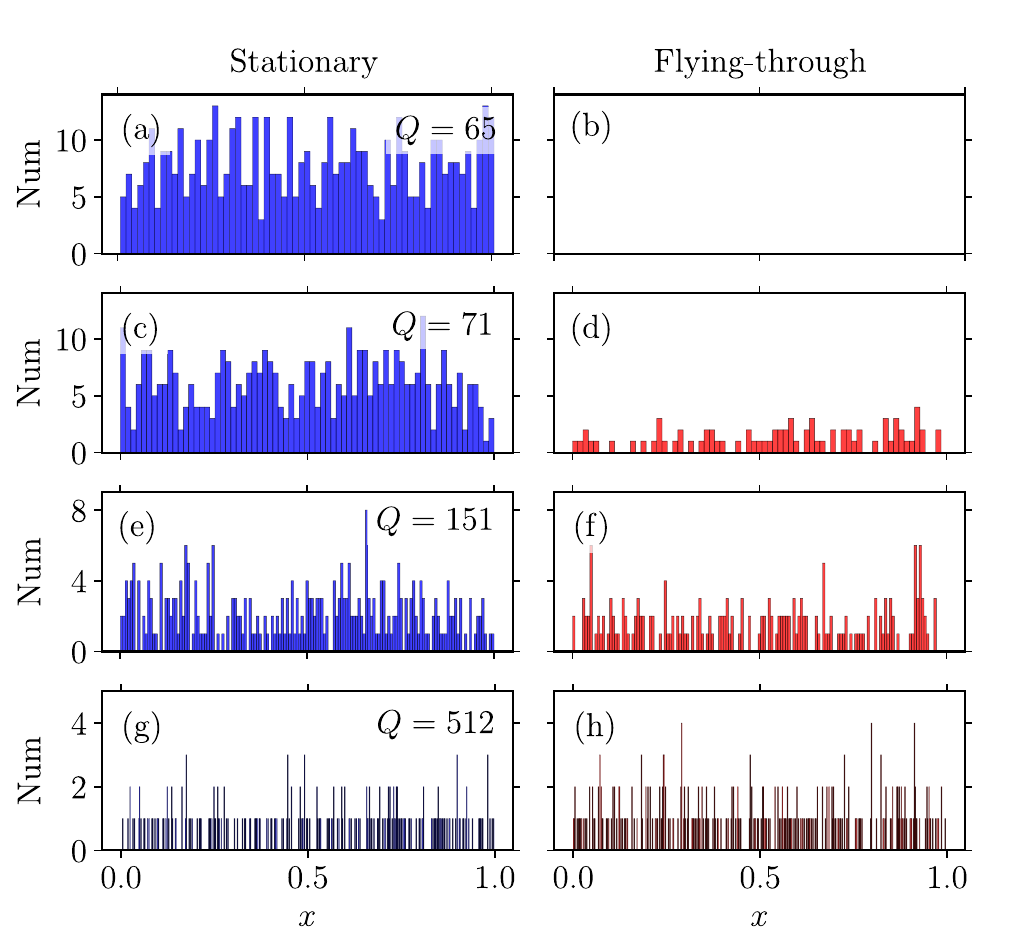}
	\caption{{Histogram of the particle number located in the vicinity of potential wells for stationary (left column) and flying-through (right column) particles for $Q=65$ (a, b), $Q=71$ (c, d), $Q=151$ (e, f) and $Q=512$ (g, h). Initial values: the initial coordinate $x_n(0)$ was chosen from a uniform distribution on the interval $[0, L]$, the initial velocity was chosen from a normal distribution $N(v_0, \sigma^2)$ with $v_0=0.0$ and $\sigma = 0.1/\sqrt{\lambda}$. Parameters: $N=512$, $L=1.0$, $\kappa=5.2$, $\mu=1.0$, $\lambda=5.0$, $\gamma=0.3$.}}
	\label{fig:1_1}
\end{figure}

In the further numerical analysis we will focus on the parameter set $\lambda = 5.0$, $\gamma=0.3$, which determines the particles' dynamics. {Fig.~\ref{fig:1_1} shows histograms of the stationary and flying-through particle number that are located in the vicinity of potential wells depending on $Q$. For $Q=65$ all particles are stationary, for $Q>Q^{*}=70$ particles are divided into subsystems of stationary and flying-through particles. It has been demonstrated that a decrease in the number of stationary particles and an increase in the number of flying-through particles with increasing parameter $Q$. For these cases we will consider phase dynamics in the following sections.}

%---------------------------------------------------------------------------%
\section{Chimera regimes in the ensemble of stationary particles}
\label{sec:stat}
Let us further consider the phase dynamics of the system~\eqref{eq:main_x}, \eqref{eq:main_phi}, \eqref{eq:ker}. For values of control parameters $\omega=0$, $\alpha=1.457$, $\kappa=5.2$ long-lived chimera states are realized in the system of phase oscillators. When the oscillators are equidistantly arranged, the fully synchronous state is also stable. As initial states for particle phases we choose $\varphi_n(0)=0$. {The initial coordinates $x_n(0)$ were sampled from uniform distribution on the segment $[0, L]$; the initial velocities $\dot{x}_n(0)$ were sampled from normal distribution $N(v_0, \sigma^2)$ with $v_0=0.0$ and $\sigma = 0.1/\sqrt{\lambda}$.} At number $Q$, not exceeding $Q^*=70$, all particles appear in potential wells, and several particles can appear in one potential well {(see Fig.~\ref{fig:1_1}a)}. Due to the inhomogeneity of the particles arrangement, the fully synchronous dynamics can become unstable, and chimera states are established in the system after the transition process (see~\cite{smirnov2021}).

To describe the degree of local phase synchronization between the elements we will use a complex local order parameter
\begin{equation}
    Z(x, t) = \langle e^{i \tilde{\varphi}} \rangle_{\textit{loc}} {=\frac{1}{|K_{\delta}(x,t)|}\sum_{k \in K_{\delta}(x,t)}e^{i \varphi_k(t)}},
    \label{eq:op}
\end{equation}
{where $K_{\delta}(x)$ is the set of particle indices located on the segment $[x-\delta, x+\delta]$}. We will separately consider the local order parameters $Z_S$ and $Z_M$ for subgroups of stationary and flying-through particles when averaging in the expression~\eqref{eq:op} is performed only for the elements from stationary or flying-through subsystems, respectively.

\begin{figure}[h!]
    \centering
    \includegraphics[width=1.0\columnwidth]{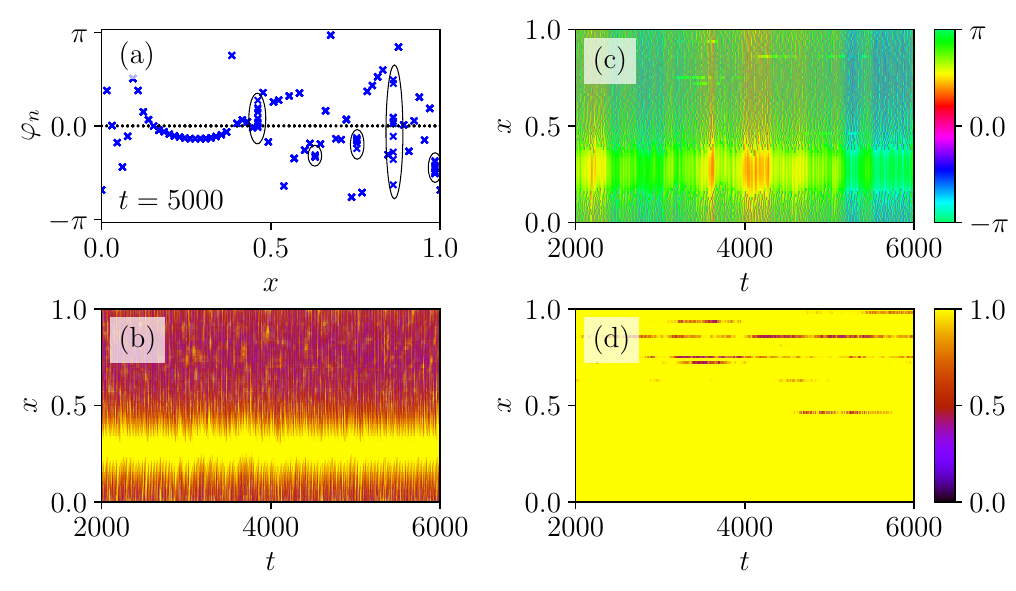}
	\caption{Chimera state in the system \eqref{eq:main_x}, \eqref{eq:main_phi}, \eqref{eq:ker} at $Q=65$. (a) Dynamics of phases $\varphi_n$, all oscillators are in potential wells (stationary). Blue cross-shaped markers -- phases of oscillators $\varphi_n$ located in potential wells, black round markers -- coordinates of potential wells. Black ellipses highlight the phases of elements that are in the same potential well but are not synchronous with each other in phase. (b) Dynamics of the local order parameter modulus for stationary oscillators $|Z_S(x, t)|$. {(c) Dynamics of $\text{arg } R(x_q,t) - \Omega t$, where $R(x_q,t) = \sum_{k \in M(x_q)}e^{i \varphi_k(t)} \Big/ |M(x_q)|$ is the modulus of the cluster order parameters, $M(x_q)$ -- the set of indices of particles located in the potential well with coordinate $x_q$, $\Omega$ -- the average rotation frequency of a fully synchronous cluster. (d) Dynamics of cluster order parameters modules $|R(x_q,t)|$.} Parameters: $N=512$, $L=1.0$, $\kappa=5.2$, $\alpha=1.457$, $\omega=0.0$, $\mu=1.0$, $\lambda=5.0$, $\gamma=0.3$.}
	\label{fig:2}
\end{figure}

In Fig.~\ref{fig:2} the chimera state for $Q=65$ is demonstrated. In this regime there is a cluster of phase-synchronous particles with $|Z(x,t)|=1$ and a region of asynchronous particles with $|Z(x,t)|<1$. Here we note that the phases of particles located in the same potential well can be incoherent (see the black curve markers in Fig.~\ref{fig:2}a).
{Figure~\ref{fig:2}d shows the order parameter modulus for each potential well. It can be seen that within the asynchronous chimera region, some clusters may become asynchronous for a long of time.}

%---------------------------------------------------------------------------%
\section{Chimera states coexistence in the ensemble of stationary and flying-through particles}
\label{sec:dyn}
\begin{figure}[h!]
    \centering
	\includegraphics[width=1.00\columnwidth]{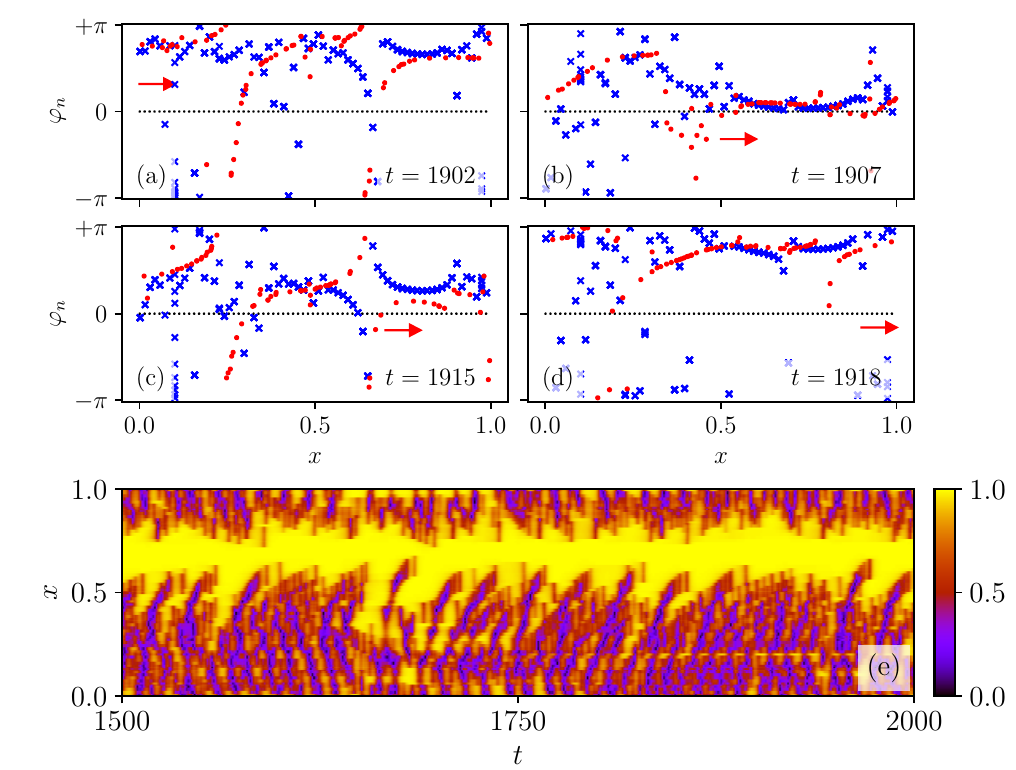}
	\caption{Chimera state in the system \eqref{eq:main_x}, \eqref{eq:main_phi}, \eqref{eq:ker} at $Q=71$. (a)--(c) Dynamics of phases $\varphi_n$ at different moments of time: (a) $t=1902$; (b) $t=1907$; (c) $t=1915$; (d) $t=1918$. Blue cross-shaped markers -- phases of oscillators $\varphi_n$ located in potential wells ($S=441$), red round markers -- phases $\varphi_m$ of flying-through particles ($M=71$), black round markers -- coordinates of potential wells. The red arrow indicates the direction of motion of flying-through particles. (e) Dynamics of the local order parameter modulus for stationary oscillators $|Z_S(x_s, t)|$. Parameters: $N=512$, $L=1.0$, $\kappa=5.2$, $\alpha=1.457$, $\omega=0.0$, $\mu=1.0$, $\lambda=5.0$, $\gamma=0.3$.}
	\label{fig:3}
\end{figure}

Let us consider the phase dynamics of the system for the case when the number of potential wells $Q$ exceeds the critical value $Q^*$. In this case, stationary and flying-through particles coexist within the same system~\eqref{eq:main_x}. When $Q=71$, a subensemble of $M=71$ flying-through particles is formed {(see Fig.~\ref{fig:1_1}c,d)}. A chimera state is also formed in the stationary particle subsystem, which is influenced by a group of flying-through particles (Fig.~\ref{fig:3}). In the moving subsystem, the behavior of the phases is relatively irregular and is largely determined by the degree of phase synchronization in the structure formed by the stationary particles. If a synchronous cluster is located there, the phases of the particles are more correlated than in the asynchronous part of the chimera. The chimera itself is also affected by running particles, its synchronous cluster can increase in size if a group of more correlated elements pass through it (see Fig.~\ref{fig:3}b). The length of the synchronous cluster fluctuates, but its location does not change on average (see Fig.~\ref{fig:3}e).

\begin{figure}[h!]
    \centering
	\includegraphics[width=1.00\columnwidth]{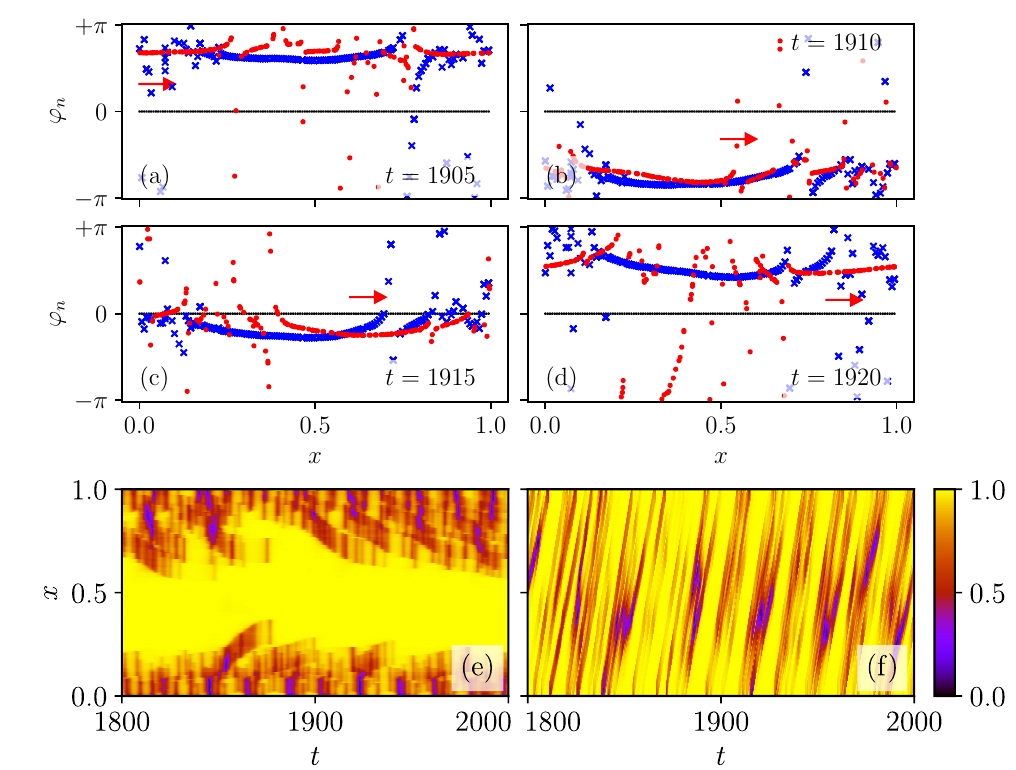}
	\caption{Stationary and flying-through chimera states coexistence in the system~\eqref{eq:main_x}, \eqref{eq:main_phi}, \eqref{eq:ker} at $Q=151$. (a)--(d) Dynamics of phases $\varphi_n$ at different moments of time: (a) $t=1905$; (b) $t=1910$; (c) $t=1915$; (d) $t=1920$. Blue cross-shaped markers -- phases of oscillators $\varphi_n$ located in potential wells ($S=327$), red round markers -- phases $\varphi_m$ of flying-through particles ($M=185$), black round markers -- coordinates of potential wells. The red arrow indicates the direction of motion of flying-through particles. (e) Dynamics of the local order parameter modulus for stationary oscillators $|Z_S(x_s, t)|$. (f) Dynamics of the local order parameter modulus for flying-through oscillators $|Z_M(x_m, t)|$. Parameters: $N=512$, $L=1.0$, $\kappa=5.2$, $\alpha=1.457$, $\omega=0.0$, $\mu=1.0$, $\lambda=5.0$, $\gamma=0.3$.}
	\label{fig:4}
\end{figure}

At $Q=151$, the subgroup of flying-through elements increases {(see Fig.~\ref{fig:1_1}e,f)} and a chimera-like running structure with a clearly distinguished synchronous domain begins to form within it (see Fig.~\ref{fig:4}). It may occasionally break up into several clusters, but their phases remain equal on average. In this case, the size of the synchronous cluster in the stationary chimera becomes larger. It can reach its largest size at moments when synchronous clusters of both structures are in the same region of space (Fig.~\ref{fig:4}b).

\begin{figure}[h!]
    \centering
	\includegraphics[width=1.00\columnwidth]{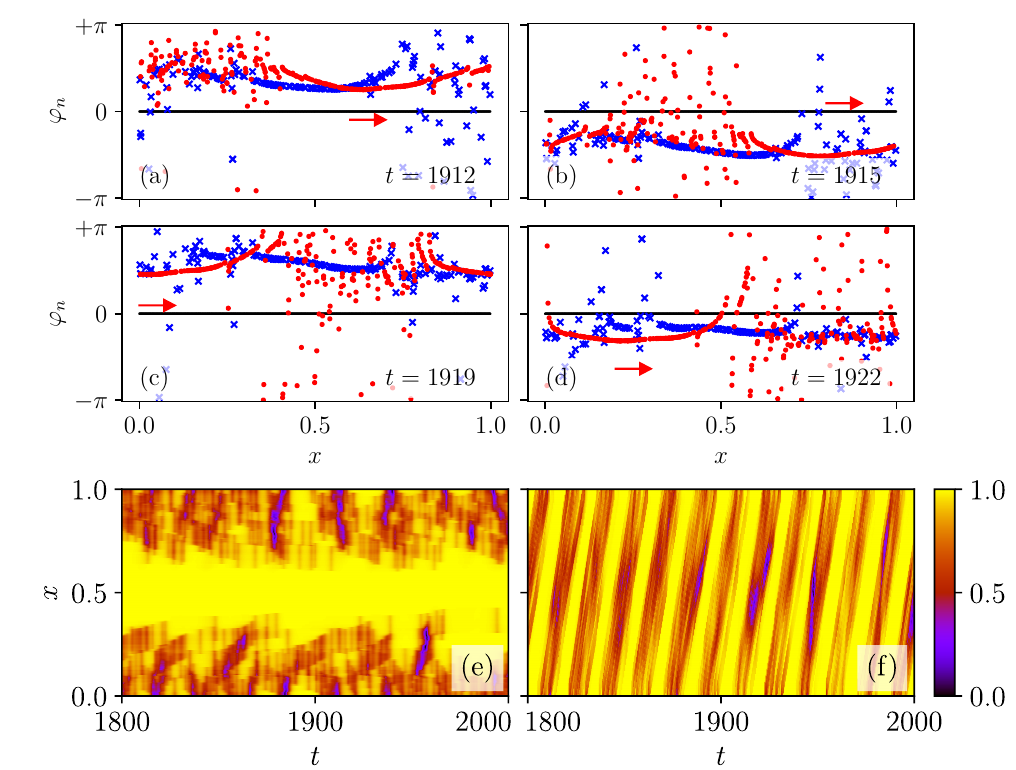}
	\caption{Stationary and flying-through chimera states coexistence in the system~\eqref{eq:main_x}, \eqref{eq:main_phi}, \eqref{eq:ker} at $Q=512$. (a)--(d) Dynamics of phases $\varphi_n$ at different moments of time: (a) $t=1905$; (b) $t=1910$; (c) $t=1915$; (d) $t=1920$. Blue cross-shaped markers -- phases of oscillators $\varphi_n$ located in potential wells ($S=198$), red round markers -- phases $\varphi_m$ of flying-through particles ($M=314$), black round markers -- coordinates of potential wells. The red arrow indicates the direction of motion of flying-through particles. (e) Dynamics of the local order parameter modulus for stationary oscillators $|Z_S(x_s, t)|$. (f) Dynamics of the local order parameter modulus for flying-through oscillators $|Z_M(x_m, t)|$. Parameters: $N=512$, $L=1.0$, $\kappa=5.2$, $\alpha=1.457$, $\omega=0.0$, $\mu=1.0$, $\lambda=5.0$, $\gamma=0.3$.}
	\label{fig:5}
\end{figure}

When $Q=512$, the number of running particles exceeds the number of stationary particles {(see Fig.~\ref{fig:1_1}g,h)}. In the running subsystem, a chimera state with a well-defined synchronous part with an almost constant drift velocity is formed (Fig.~\ref{fig:5}). The size of the synchronous cluster in the subensemble of stationary particles becomes smaller due to the decrease in the number of corresponding elements.

Thus, an atypical effect of coexistence of chimera states in two subsystems of stationary and flying-through particles whose phases are nonlocally coupled is observed. Despite the fact that the mean field formed by all elements is nonstationary due to the presence of running particles, a quasi-static (in a stationary or uniformly moving coordinate system) chimera structure is formed in each of the subsystems.
%---------------------------------------------------------------------------%
\clearpage
\section{Conclusion}\label{sec:Conclusion}
{
This work greatly extends our understanding of the dynamics of flying-through oscillators in media with nonlocal interaction. In our previous work~\cite{smirnov2021} it was shown that in the case of non-equidistant arrangement of particles along the medium or at diffusive or ballistic motion of oscillators the fully synchronous state can become unstable and the chimera state becomes the dominating regimes. Here we assume that the particles are in a periodic potential under the influence of an external force and dissipation.
}

{
Let's describe the results obtained. First, it was shown that, depending on the number of potential wells, particles can be stationary, flying-through, or split into two subensembles of stationary and flying-through particles. Secondly, it has been demonstrated that in the case of stationary arrangement of particles, the fully synchronous regime can become unstable due to the inhomogeneity of particle distribution between potential wells. In a partially synchronous cluster, phase incoherent elements can be located within the same potential well. Third, when stationary and flying-through elements coexist, chimera regimes are realized at each subensemble and further coexist with each other. This means that, despite the nonlocal nature of the interaction between elements of the medium, what is important for structure formation is a relatively stable mutual arrangement of particles relative to each other, which leads to a chimera state in each of the subensemble.
}
%---------------------------------------------------------------------------%
\section*{Acknowledgments}
{
We thank Arkady Pikovsky for useful discussion and comments. This work was supported by the Ministry of Science and
Higher Education of the Russian Federation under project No.
FSWR-2020-0036 (Sec.~\ref{sec:model}), the Russian Science Foundation under project No. 23-12-00180 (Sec.~\ref{sec:stat},~\ref{sec:dyn}).
}
%-----------------------------------------------------------------
\section*{Data availability}
The data that support the findings of this study are available from the corresponding author upon reasonable request.
%---------------------------------------------------------------------------%
\bibliography{references}
%---------------------------------------------------------------------------%
\end{document}